\documentclass[a4paper]{statsoc}

\usepackage[T1]{fontenc}
\usepackage[utf8]{inputenc}

\usepackage{fullpage}
\usepackage{natbib}

\usepackage{amscd,amsmath,latexsym,amssymb,amsfonts}
\usepackage{mathrsfs} 

\usepackage{bm}

\usepackage[a4paper,left=2cm,top=2cm,right=2cm,bottom=2cm,ignoreheadfoot]{geometry}

\title{A discussion of the paper ``Safe testing" by Gr\"unwald, de Heide, and Koolen, Read before The Royal Statistical Society at a meeting organized by the Research Section on Wednesday, 24 January, 2024}
\author{Joshua Bon$^{1}$ and Christian P.  Robert$^{1,2}$}
\address{$^{1}$Université Paris-Dauphine PSL and $^2$University of Warwick}

\begin{document}
\maketitle 

\begin{abstract}
This is a discussion of the paper ``Safe testing" by Gr\"unwald, de Heide, and Koolen, Read before The Royal Statistical Society at a meeting organized by the Research Section on Wednesday, 24 January, 2024.
\end{abstract}

The central proposal found in this quite original paper could prove relevant for the so-called {\em objective Bayes} approach \citep{berger:1985,robert:2001} as
it may open a way to deriving prior distributions from (frequentist) first principles. While some Bayesian schools refuse to consider the concept of testing the veracity of a null hypothesis, most do propose approaches and methodologies to address this challenge.  These mainly rely on Jeffreys' formalisation via two prior distributions \citep{Wrinch:Jeffreys:1919,jeffreys:1939} via what would later come to be the Bayes Factor \citep{robert:chopin:rousseau:2009,ly:verhagen:wagenmakers:2016,wagenmakers:ly:2022}. In his {\em Theory of Probability}, Jeffreys constructs the equivalent of his point estimation prior for tests, although it had little impact on the field until rediscovered by \cite{bayarri:garcia:2007}.

We commend the points put forward to support $e$-values,
\begin{enumerate}
\item behaviour under optional continuation [by a martingale reasoning]

\item interpretation as `evidence against the null' as gambling

\item in all cases preserving frequentist Type I error guarantees

\item $e$-variables turn out to be Bayes factors based on the right Haar prior (rather than at times on highly unusual (e.g. degenerate) priors?)

\item $e$-variables need more extreme data than $p$-values in order to reject the null
\end{enumerate}
since they are mostly worthwhile, even though (b) is vague and (c) is frequentist. The link of the theory of $e$-values with right Haar priors reminds us of Pitman's derivation of the best equivariant estimator \citep{robert:2001}, even though the fact that Haar priors are improper for the most part proves a major hindrance in a testing situation \citep{degroot:1973}. The optional continuation in (a) is an even stronger argument---if from a Bayesian viewpoint---since it is been used to defend the Bayesian paradigm. On the other hand, if sequentially testing e-values, only the first e-value will be interpretable as a coherent Bayes factor. That is, in sequential testing, the product of the e-values does not share the same properties as a product of Bayes factors (i.e. cancellation of terms as a telescoping product). 

Central to our interest, point (d)~above, despite the cautionary added label of an `Almost Bayesian Case', brings a formal way to define {\em least favourable priors} \citep{berger:1985} in a testing environment, albeit in no clear connection with the reference priors of \cite{bayarri:garcia:2007}. This derivation, formalised as Theorem 1, provides a reversal of the typical prior choice in testing. A more common stance is to hold the (subjectively chosen) prior distribution on $H_0$ as a starting point---this is an obviously known entity. One may wonder if a dual perspective, i.e., when starting from the prior distribution on $H_0$, would prove fruitful, e.g., in leading to an optimal prior on $H_1$. 

The paper seems to conclude at a perfect duality with the minimax-maximin result. This also relates to the projection priors we defined in \cite{goutis:robert:1998} and \cite{dupuis:robert:2003}. 

\medskip
\noindent {\bf Example 1.} In order to examine this potential connection, let us examine the formal situation when $Y\sim\mathcal N(\mu,1)$ and the null hypothesis is$$H_0:\ \mu\in(-a,a)$$as in \cite{casella:strawderman:1981}, selecting for our prior under $H_1$ a conjugate Normal $\mu\sim\mathcal N(0,1)$. The marginal $m(y)$ is then centred at zero and this seems to imply that the optimal prior on $H_0$ is a point mass at either extremity (or maybe balanced over both), whatever the value of $a$. Contrary to the point estimation case \citep{casella:strawderman:1981}, where the minimax prior is allowing for a number of support points increasing with $a$, this solution is very rudimentary and hence of little utility.

\medskip
As a plus, however, the approach found therein allows for improper priors on the nuisance parameters and thus somehow brings a stronger justification than \cite{berger:pericchi:varshavsky:1998} in favour of the permanence of an identical measure (on the nuisance parameters) across hypotheses and models.

In conclusion, we congratulate the authors on this endeavour but it remains unclear to us (as Bayesians) (i) how to construct the least favorable prior on $H_0$ on a general basis, especially from a computational viewpoint, and, more importantly, (ii)  whether it is at all of inferential interest [i.e., whether it degenerates into a point mass]. 
With respect to the sequential directions of the paper, we also wonder at the potential connections with sequential Monte Carlo, for instance, towards conducting sequential model choice by constructing efficiently an amalgamated evidence value when the product of Bayes factors is not a Bayes factor \citep[see, e.g.][]{buchholz:etal:2022}.

\section*{Acknowledgement}
Funded by a Prairie chair from the Agence Nationale de la Recherche (ANR-19-P3IA-0001) and by the European Union (ERC-2022-SYG-OCEAN-101071601).
Views and opinions expressed are however those of the author(s) only and do not necessarily reflect those of the a Prairie chair from the Agence Nationale de la Recherche (ANR-19-P3IA-0001), the European Union or the European Research Council Executive Agency. Neither the European Union nor the granting authority can be held responsible for them.

\bibliographystyle{plainnat}
\bibliography{biblio}
\end{document}